\documentclass[showpacs,preprintnumbers,amsmath,amssymb]{revtex4}

\usepackage{graphicx}
\usepackage{dcolumn}
\usepackage{bm}

\begin{document}


\title{Correlated dynamics of weakly charged silica spheres at an air-water interface\footnote{Corresponding author Tel.: +86 21 55665338}
}

\author{Wei Zhang$^{1,2}$}

\author{Wei Chen$^{1}$}%
 \email{phchenwei@gmail.com}

\author{Penger Tong$^{3}$}%

\affiliation{%
$^1$Department of Physics, Fudan University, Shanghai 200433, China\\
$^2$Department of Physics, Jinan University, Guangzhou 510632, China\\
$^3$Department of Physics, Hong Kong University of Science and Technology, Clear Water Bay, Howloon, Hong Kong}

\date{\today}

\begin{abstract}
Abstract - Optical microscopy and multi-particle tracking are used to investigate the spatially correlated motion of weakly charged silica spheres at an air-water interface for different area fraction $n$ occupied by the particles. When the area fraction is very small, e.g. $n=0.03$, the correlation function along the line joining the centers of particles $D_{rr}$ decays with inter-particle distance $R$ as $1/R^{0.86\pm0.02}$, and the function perpendicular to this line $D_{\theta\theta}$ decays with $R$ as $1/R^{1.45\pm0.03}$, which differs from the results of [Phys. Rev. Lett. 97, 176001 (2006)] with low surface viscosity (where $D_{rr}\propto 1/R$, $D_{\theta\theta}\propto 1/R^2$). We argue that the differences arise from the Coulomb interaction between particles. The Coulomb interaction enhances the correlated motion of particles. Experimental results show that with the increase of $n$, the decay rate of $D_{rr}$ and $D_{\theta\theta}$ with $R$ decreases and the cross-correlation enhances for the Coulomb interaction increases. The Coulomb interaction between colloidal particles should serve as an effective surface viscoelastical role in our system. With the scaled separation $\frac R {d}\left(\frac {\eta_{w}d} {\eta_{es,2p}}\right)^{3/2}$, the correlated motions for various values of $n$ and different particles can be scaled onto a single master curve, where $d$ is particles' diameter, $\eta_{w}$ is the viscosity of the water,  and $\eta_{es,2p}$ is the effective surface viscosity whose measurements agree well with that of one-particle surface viscosity $\eta_{es,1p}$. The effective surface viscosity $\eta_{es,2p}$ as a function of the area fraction $n$ for different silica spheres is presented.

\end{abstract}

\pacs{82.70.Dd, 68.05.Gh, 05.40.-a, 83.85.Jn}
\maketitle

\section{\label{sec:level1}Introduction}

Colloidal particles moving through a fluid create a flow that affects the movement of other particles in its vicinity. This interaction between colloidal particles is referred to as hydrodynamic interaction. The hydrodynamic interaction of bulk suspensions is long-ranged and decays with inter-particle distance $R$ as $1/R$ \cite{1,2}. In many real circumstances, particles are not completely free, but spatially confined in some special environments, such as microfluidic devices, porous media, fluid interface and cell membrane\cite{3,4,5}. The hydrodynamics behaviors of confined colloidal suspensions are one of the current investigation focus. In 2000, E. R. Dufresne $et$ $al.$ \cite{6} measured the dynamic coupling of particle pair above a single flat wall using optical tweezers. Their experimental results showed that while the distance of particles to the wall is about three times of particle radius, the hydrodynamics interaction between particle pair decays with $R$ as $1/R^2$. In 2011, P. L. Pushkar $et$ $al.$ \cite{7} made a further investigation on many-body hydrodynamic interaction in the presence of a single wall by holographic optical tweezers and dynamic simulation. They found that the behaviors of hydrodynamic interactions between particle pair and multi-particle clusters are similar, i.e., the hydrodynamics interaction between particles in clusters also decays with $R$ as $1/R^2$. This results differ from that in bulk in which hydrodynamic screen appears in high particle concentration. B. Cui $et$ $al.$ investigated the hydrodynamic interaction between particles in a narrow channel \cite{7} and confined between two plates \cite{9} in 2002 and 2004, respectively. They found that a linear channel sharply screens the hydrodynamic interaction; and the hydrodynamic interaction of particles confined between two plane walls decays with $R$ as $1/R^2$. Perpendicular to the line joining the centers of particles confined between two plates the cross-correlation is negative, i.e., particles exert anti-drag on one another in this direction. The circulation current is the direct cause for this negative coupling. The above works revealed the corrected dynamic characters of colloidal particles confined by hard wall.

There is a non-slip boundary condition near a hard wall which could cut off the fluid field in its vicinity. The fluid interface is a slip boundary and may not cut off the fluid field. The influences of boundary condition reflect directly on the correlated dynamics of particles. Because of the effects of the spatial symmetry breaking, the hydrodynamic interaction of particles at an air-water interface is different from that in symmetrical bulk. In 2006, V. Prasad $et$ $al.$ \cite{10} investigated the hydrodynamic interaction between particles at an air-water interface and found that there is a transition from three-dimension (3D) to two-dimension (2D) hydrodynamics with increasing surface viscosity $\eta_{s}$. At high $\eta_{s}$, the correlated motion along the line joining the centers of particles $D_{rr}$, and ones perpendicular to this line $D_{\theta\theta}$ are nearly equal and have logarithmic $R$ dependence. With the decrease of $\eta_{s}$, the decay rate of $D_{rr}$ and $D_{\theta\theta}$ with $R$ increases and the their values begin to deviate. At low $\eta_{s}$, $D_{rr}\propto 1/R$, $D_{\theta\theta}\propto 1/R^2$. Notwithstanding the differences in the behavior of $D_{rr}$, $D_{\theta\theta}$ at high and low $\eta_{s}$, all the data sets can be scaled onto a single master curve by $\eta_{s}$. Evidently, the surface viscosity $\eta_{s}$ has a substantial influence on both $D_{rr}$ and $D_{\theta\theta}$. In 2011, M. H. Lee $et$ $al.$ \cite{11} investigated the correlated motion of particles at an oil-water interface. Their experiment results also show a transformation from 3D to 2D hydrodynamics with the increase of $\eta_{s}$. The results of Prasad $et$ $al.$ have the effect of fluid field only from one side; and ones of Lee $et$ $al.$ have the influence of the fluid field from two side. The hydrodynamic interaction between particles at an oil-water interface is influenced by the viscosity of the oil $\eta_{oil}$, water $\eta_{w}$ and interface $\eta_{int}$. From the above, we know that the dynamic characters of colloidal system in the confined condition exhibited much more rich physical image than that in the bulk. The investigations on the dynamics of confined colloidal suspensions have significant application in many important field, such as pharmacy and Biological science etc. \cite{12}.

We are interested in the influence of the hydrodynamic and Coulomb interactions on dynamic characters of colloidal system when the boundary condition changes from the symmetry to completely asymmetry, such as air-water interface. In the previous experiment investigations about the hydrodynamic interactions between particles at an air(oil)-water interface the interface viscosity $\eta_{int}$ was treated as an adjustment parameters. In this paper, we measured the cross correlation of colloidal interfacial diffusion under the different interfacial concentrations of weakly charged colloidal spheres, and investigate the influence of Coulomb interaction between the spheres on the correlated motion.

Getting rid of the impurity from the aqueous surface, we make a clean air-water interface and investigate the correlated motions of silica spheres S1 ($d=1.57\mu m$), S2 ($d=0.97\mu m$) and S3 ($d=0.73\mu m$) at this pure interface by optical microscopy and multi-particle tracking. The experiment results show that the cross correlation of particles at pure air-water interface is stronger than that exhibited in the experiments of Prasad $et$ $al.$ \cite{10} in the case of low $\eta_{s}$. The correlated motion $D_{rr}\propto 1/R^{0.86\pm0.02}$ turns into $D_{rr}\propto 1/R^{0.68\pm0.02}$ when $n$ was from 0.03 to 0.35. The particles used in our experiment are weakly charged silica spheres. Coulomb interactions between particles enhanced the cross correlation of particles. With the increase of particles' area fraction $n$, the mean particle separation decreases and Coulomb interactions increase. Thus the cross correlation of particles enhances and the decay rate of $D_{rr}$ and $D_{\theta\theta}$ with $R$ decreases as $n$ is increased. The correlated motion of particles as a function of inter-particle distance $R$ for different $n$ is presented. As discussed in the paper, the Coulomb interaction between colloidal particles should serve as an effective surface viscoelastical role in our system. Though the differences in the behavior of correlated motion for different $n$ and particles, with the scaled separation $\frac R {d}\left(\frac {\eta_{w}d} {\eta_{es,2p}}\right)^{3/2}$ and correlation function $<D_{rr,\theta\theta}/\tau>/2D_s$, all the correlated motion can be scaled onto a single master curve. Where $D_s=k_BT/4\pi\eta_{es,2p}$ and $\eta_{es,2p}$ is termed as two-particle effective surface viscosity whose measurements agree well with that of one-particle surface viscosity $\eta_{es,1p}$. We present the effective surface viscosity $\eta_{es,2p}$ as a function of the area fraction $n$ for different silica spheres. Particles S3 exhibit an anomalous long-range correlation which we do not understand at present.

The paper is organized as follows. Section II is devoted to the description of the experiment setup and results. In section III, data analysis are presented. In section IV, further discussions are given. Finally, the work is summarized in section V.

\section{Experiment setup and results}

The particles used in the experiment are silica sphere 1 (S1)with the diameter $d=1.57\pm 0.06\mu m$, silica sphere 2 (S2) with the diameter $d=0.97\pm 0.05\mu m$ and silica sphere 3 (S3) with the diameter $d=0.73\pm 0.04\mu m$. The S1 and S3 particles were purchased from Duke Scientific and were synthesized by the St\"{o}ber method, followed by a sintering process at $900^0C$ to drive off all organic materials and water. The particles S2 were purchased from Bangs Laboratories and are non porous spheres. We follow the same procedures as described in Ref.[13] for the experimental setup and the sample preparation and cleaning. The sample cell is made from a stainless steel disk (which is hydrophilic) with a central hole of diameter 13mm. The bottom of the hole is sealed with a 0.1mm thick glass cover slip, which also serves as an optical window. The sidewall of the hole together with the bottom glass slip forms a container, which has an effective height of 1.0mm. The top of the container has a sharp circular edge, which is used to pin the air-water interface and reduce unwanted drift. The entire cell is placed on the sample stage of an inverted microscope, so that the motion of the interfacial particles can be viewed from below and recorded by a digital camera.

We take great care to clean the particle samples and the air-water interface. Surface pressure measurements reveal that no detectable impurity is found in the cleaned particle samples. The particle-methanol solution is then injected onto a cleaned interface using a syringe pump. The silica spheres disperse well at the clean air-water interface. Individual particles undergo vigorous Brownian motion and remain stable at the interface for days. These particles remain in focus under high magnification, indicating that the silica spheres are strongly bound to the interface and their vertical position is determined by an energy minimum. The usual capillary effect is not applicable to micron-sized spheres for the gravitational energy of the particles is much smaller than the relevant energy of the interface, and particles can choose an equilibrium vertical position without introducing long-range deformations to the interface.

To obtain good images of the silica particles phase contrast microscopy is used. Using homemade software with a spatial resolution of 60-100nm we construct particle trajectories from the consecutive images. From the particle trajectories, we determine the vector displacements of the tracers $\Delta r(t,\tau)=r(t+\tau)-r(t)$, where $t$ is the absolute time and $\tau$ is the lag time. In the experiment, we measure the mean cross-correlated particle motion [2]:

 \[
D_{\alpha \beta}(r,\tau)=<\Delta r^i_\alpha(t,\tau)\Delta r^j_\beta(t,\tau)\delta(r-R^{i,j}(t))>_{i\neq j},\qquad\qquad  (1)
\]
where $i,j$ are particle indicates, $\alpha$ and $\beta$ represent different coordinates, and $R^{i,j}$ is the distance between particle $i$ and $j$. The off-diagonal elements are assumed to be uncorrelated. We focus on the diagonal elements of this tensor product: $D_{rr}$ which indicates the correlated motion along the line joining the centers of particles, and $D_{\theta\theta}$ which represents the correlated motion perpendicular to this line.

The correlated motion $D_{rr}$ and $D_{\theta\theta}$ are the function of $R$ and $\tau$. The lag time $\tau$ need approach to zero to eliminate its effect on $R=R(\tau)$. There are two ways to calculate the $\tau$-independent quantities $<D_{rr}/\tau>_{\tau}$ and $<D_{\theta\theta}/\tau>_{\tau}$, which depend only on $R$. Firstly, if the data shows $D_{rr}$, $D_{\theta\theta}\propto \tau$ within the observation period of $\tau$, the slopes of $D_{rr}$, $D_{\theta\theta}$ versus $\tau$ are given by $D_{rr}/\tau$ and $D_{\theta\theta}/\tau$. Otherwise, we calculate the partial differential of $\partial D_{rr}/\partial \tau$ and $\partial D_{\theta\theta}/\partial \tau$ from which we derived the corresponding slope value at $\tau=0$ by the extrapolation.

We measured the curve of $D_{rr}/\tau$ and $D_{\theta\theta}/\tau$ vs. $R$ for dozens of different area fraction in range of 0.01-0.35 for the data of S1. We make a coarse grain process by grouping the dozen of curves to range: 0.01-0.05, 0.05-0.1, 0.1-0.14, 0.15-0.17, 0.2, 0.35 (see the figures below). We average curves in each groups respectively. The similar processing procedure had also been down for the data of S2 and S3. The results are obtained by averaging over $10^6$ particles, ensuring that the statistical averaging is adequate.

Figure 1 exhibits the measured values of $<D_{rr}/\tau>$ and $<D_{\theta\theta}/\tau>$ for silica microspheres S1 scaled by $D_0=k_BT/3\pi \eta_w d$ as a function of $R/d$ with the area fraction $n$ ranging from $0.01$ to $0.05$, where $\eta_w$ is the bulk viscosity of water in room temperature. The average result of these curves were drawn in the inset of Fig.1, while the corresponding area fraction $n$ was estimated as $0.03$. Solid and open symbols represent $<D_{rr}/\tau>/D_0$ and $<D_{\theta\theta}/\tau>/D_0$, respectively. The error bar in the inset of Fig. 1 indicates only the error range of the values of correlated motion, but it's slope. With the increase of inter-particle distance $R$, $<D_{rr}/\tau>$ and $<D_{\theta\theta}/\tau>$ decrease following the power law. As shown in Fig. 1, $<D_{rr}/\tau>\propto 1/R^{0.86\pm0.02}$ and $<D_{\theta\theta}/\tau>\propto 1/R^{1.45\pm0.03}$ which differs from experiment results of Prasad $et$ $al.$ \cite{10} with low surface viscosity where $D_{rr}$ $\varpropto$ $1/R$, $D_{\theta\theta}$ $\varpropto$ $1/R^2$.

Figure 2 (a) exhibits the measured values of $<D_{rr}/\tau>$ and $<D_{\theta\theta}/\tau>$ scaled by $D_0$ as a function of $R/d$ with various estimated values of area fraction $n$ for silica microspheres S1. Different scattering symbol represents the data with different area fraction of particles. As shown in Fig. 2 (a), with the increase of area fraction $n$, the magnitude of curves of $<D_{rr}/\tau>$ and $<D_{\theta\theta}/\tau>$ increase. We fit the data in Fig. 2(a) by $<D_{rr}/\tau>\propto 1/R^{\lambda_r}$ and $<D_{\theta\theta}/\tau>\propto 1/R^{\lambda_{\theta}}$. Figure 2 (b) shows the fitted parameters $\lambda_r$ and $\lambda_{\theta}$, which characters the decay rate of cross correlation between particles, as a function of $n$. As shown in Fig. 2 (b), with the increase of $n$, $\lambda_r$ and $\lambda_{\theta}$ decrease indicating that the decay rate of correlated motion vs. $R$ decreases. Thus, the cross correlation of colloidal diffusion was enhanced with increasing area fraction $n$.

Figure 3 (a) and (b) shows the measured values of $<D_{rr}/\tau>/D_0$ and $<D_{\theta\theta}/\tau>/D_0$ vs. $R/d$ with various values of area fraction $n$ for silica microspheres S2 and S3, respectively. The data processing for S2 and S3 is similar to that of S1. Every curve is derived by averaging the data with adjacent area fraction. The data in Fig.3 (a) are qualitatively identical with that in Fig. 2 (a): the correlated motion decays with the inter-particle distance $R$ following the power law. As the area fraction $n$ is increased, the curves shift to higher values and the decay rate of $<D_{rr}/\tau>$ and $<D_{\theta\theta}/\tau>$ versus $R$ becomes slow.

Figure 3 (b) shows an anomalous long-rang correlation. For S3 particles, when $n\geq 0.08$ and $R>5d$, the decay of $<D_{rr}/\tau>$ and $<D_{\theta\theta}/\tau>$ becomes slow with increasing $R$ and does not follow power law. At large $R$ region, $<D_{rr}/\tau>$ and $<D_{\theta\theta}/\tau>$ are weakly dependent on inter-particle distance $R$. For short inter-particle distance ($R<5d$), the behaviors of $<D_{rr}/\tau>$ and $<D_{\theta\theta}/\tau>$ are similar to that of S1 and S2 particles.

\begin{figure}
\includegraphics[height=7cm]{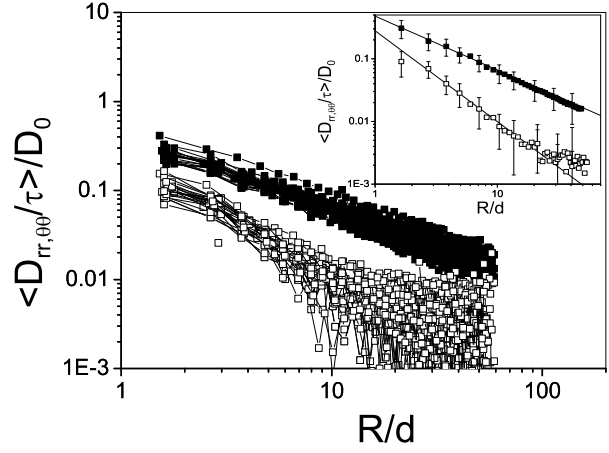}
\caption{\label{fig:epsart} Correlation function $<D_{rr}/\tau>$ (solid symbols) and $<D_{\theta\theta}/\tau>$ (open symbols) (scaled by $D_0$) as a function of inter-particle distance $R$ (scaled by $d$) with the area fraction $n$ ranging from $0.01$ to $0.05$. The average result of these curves were drown in the inset, while the corresponding area fraction $n$ was estimated as $0.03$. The solid lines in the inset are the fitted line $<D_{rr}/\tau>=0.48/R^{0.86\pm0.02}$ and $<D_{\theta\theta}/\tau>=0.28/R^{1.45\pm0.03}$. The samples are S1 silica $d=1.57\pm 0.06\mu m$.}
\end{figure}

\begin{figure}
\includegraphics[height=6cm]{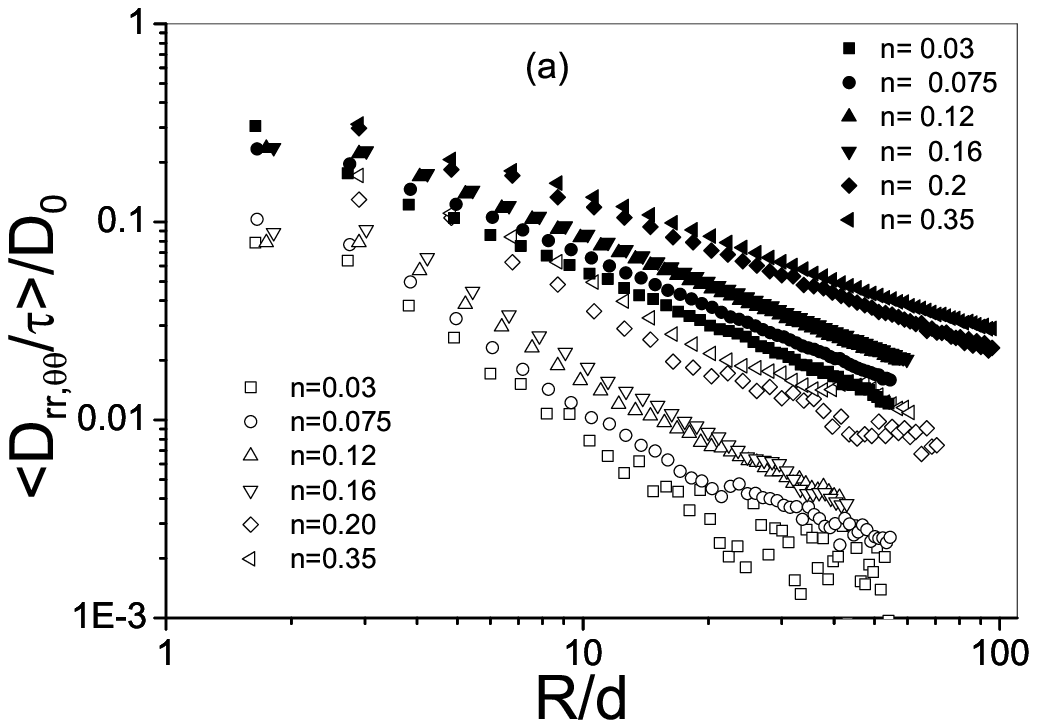}
\includegraphics[height=6cm]{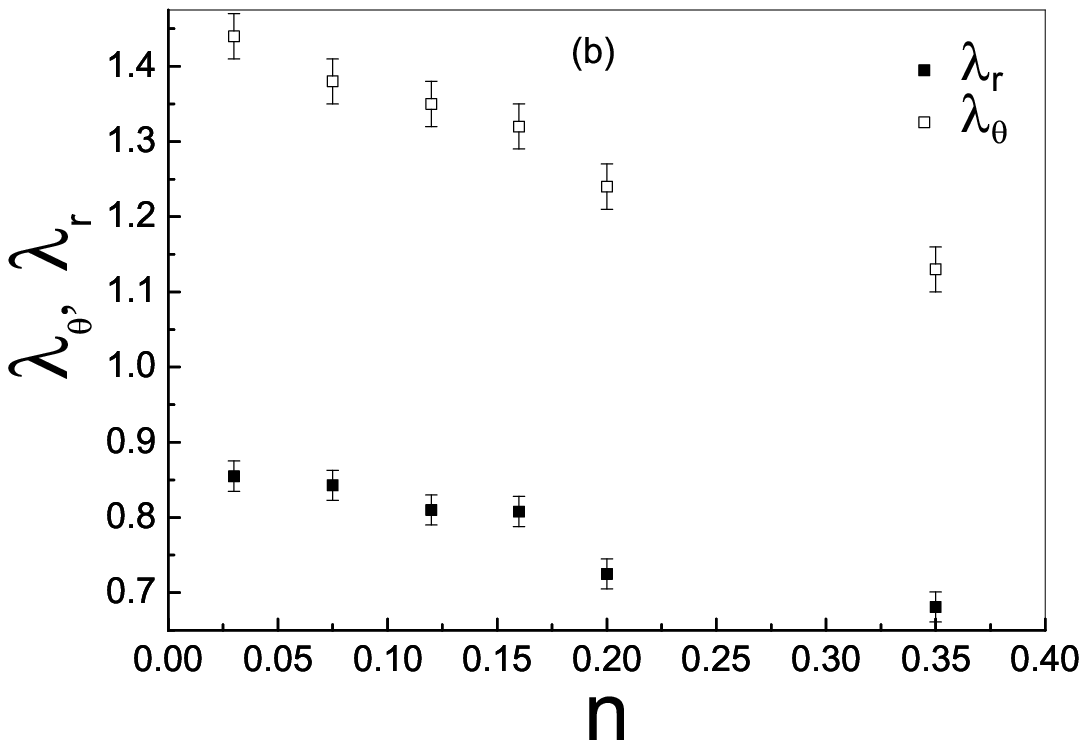}
\caption{\label{fig:epsart} (a) Correlation function $<D_{rr}/\tau>$ (solid symbols) and $<D_{\theta\theta}/\tau>$ (open symbols) (scaled by $D_0$) as a function of inter-particle distance $R$ (scaled by $d$) with various values of area fraction $n$. Different symbols represent the data for different area fraction $n$. The data in the figure were fitted by $<D_{rr}/\tau>\propto 1/R^{\lambda_r}$ and $<D_{\theta\theta}/\tau>\propto 1/R^{\lambda_{\theta}}$. (b) The fitted parameter $\lambda_{r}$ and $\lambda_{\theta}$ as a function of the area fraction $n$ for S1 silica $d=1.57\pm 0.06\mu m$. The samples are S1 silica $d=1.57\pm 0.06\mu m$.}
\end{figure}

\begin{figure}
\includegraphics[height=6cm]{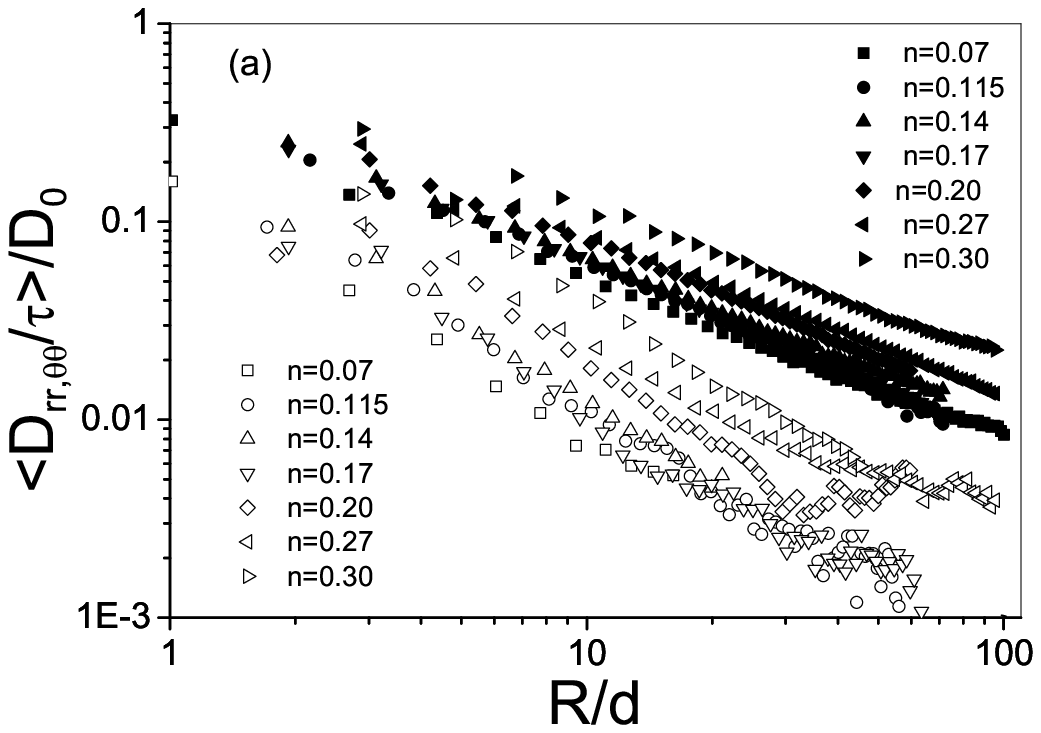}
\includegraphics[height=6cm]{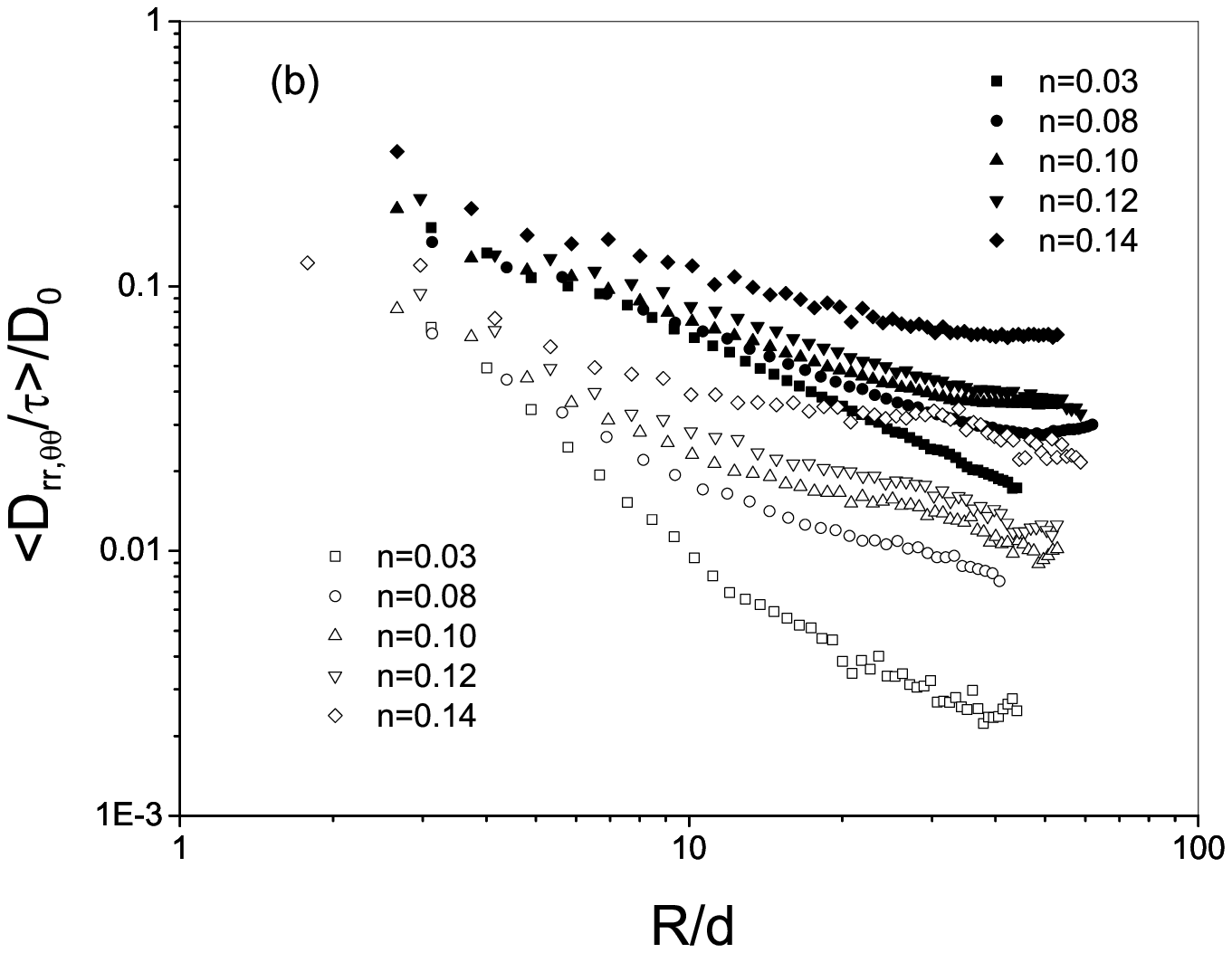}
\caption{\label{fig:epsart} Correlation function $<D_{rr}/\tau>$ (solid symbols) and $<D_{\theta\theta}/\tau>$ (open symbols) (scaled by $D_0$) as a function of inter-particle distance $R$ (scaled by $d$) with various values of area fraction $n$. Different symbols represent the data for different area fraction $n$. The samples are the following: (a) S2 silica $d=0.97\pm 0.05\mu m$; (b) S3 silica $d=0.73\pm 0.04\mu m$.}
\end{figure}

\section{Data analysis}

\begin{figure}
\includegraphics[height=6cm]{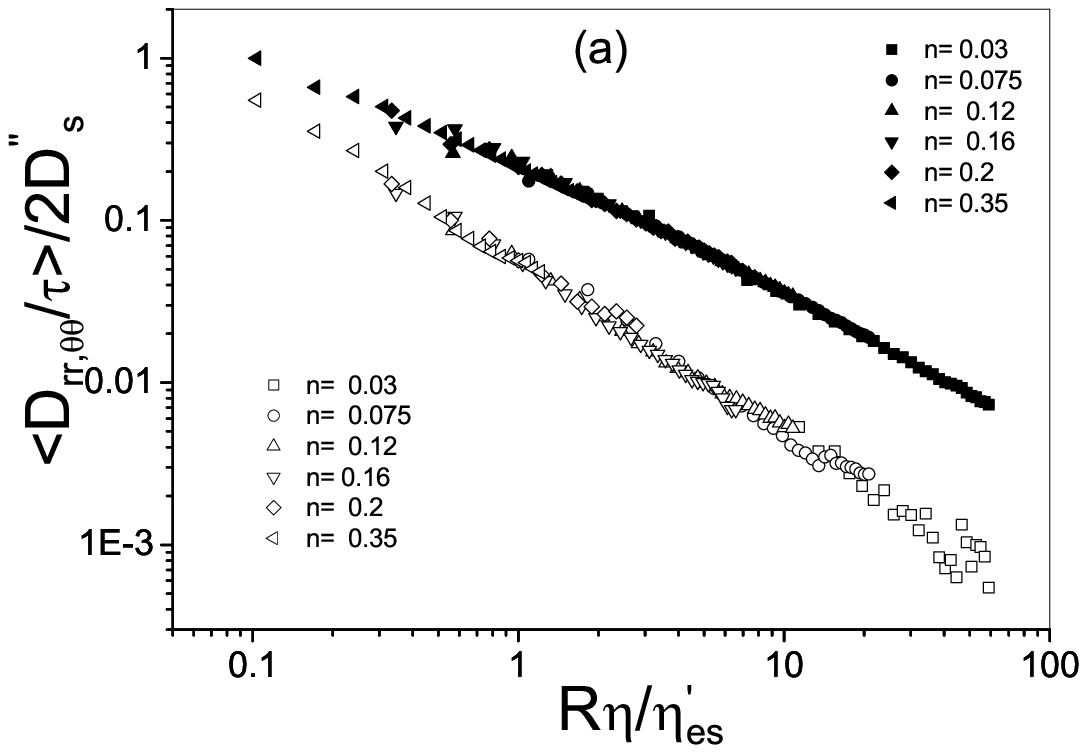}
\end{figure}

\begin{figure}
\includegraphics[height=6cm]{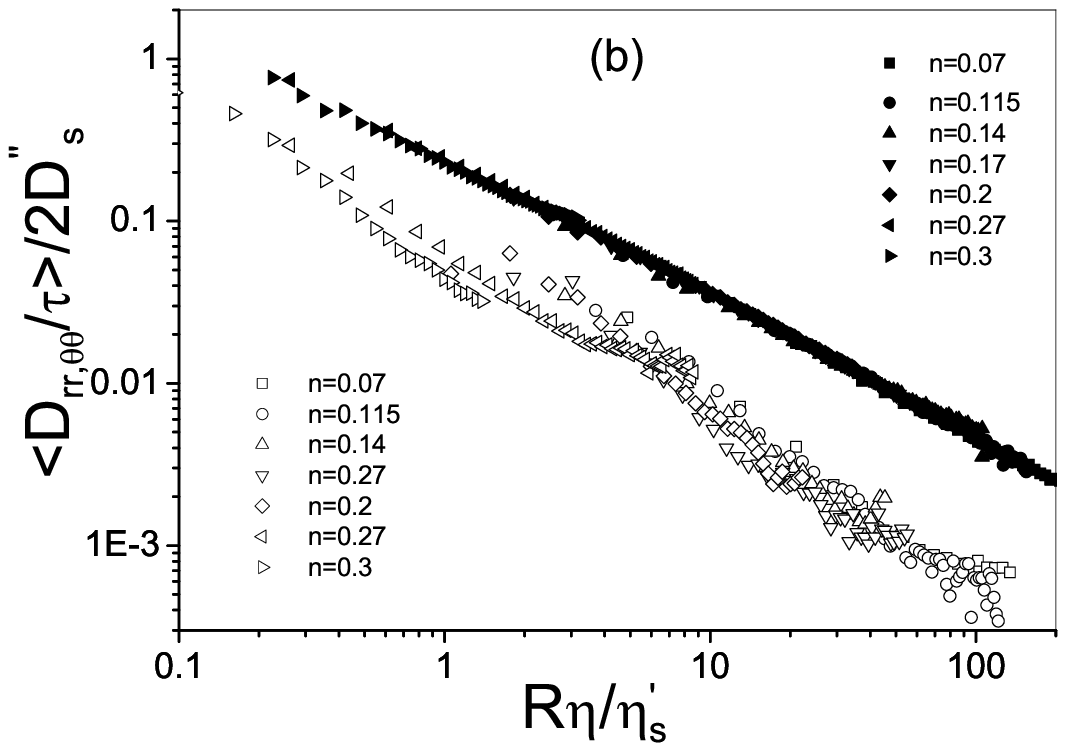}
\includegraphics[height=6cm]{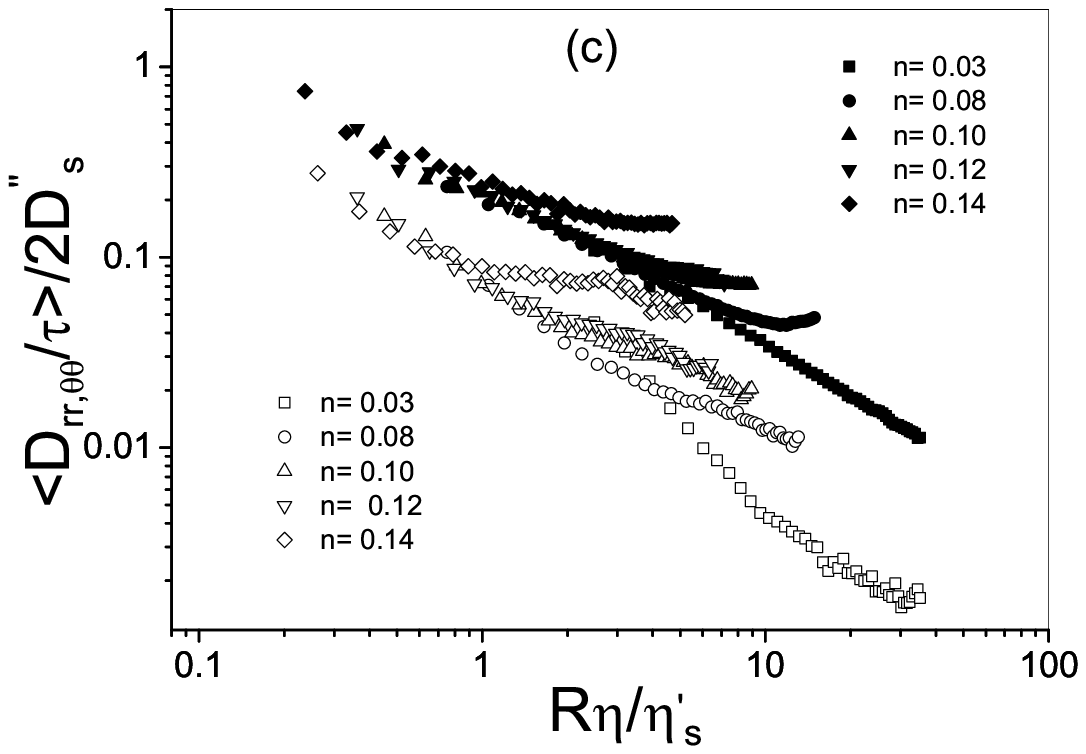}
\caption{\label{fig:epsart} Master curve of the correlation function $<D_{rr}/\tau>$ (solid symbols) and $<D_{\theta\theta}/\tau>$ (open symbols) (scaled by $2D^{''}_s$) as a function of reduced particle separation $S=R\eta_w / \eta^{'}_s$ with various values of area fraction $n$. Different symbols represent the data for different area fraction $n$. The data and symbols in Fig. 4 (a), (b) and (c) are same as that in Fig. 2 (a), 3 (a) and (b), respectively. The samples are the following: (a) S1 silica $d=1.57\pm 0.06\mu m$; (b) S2 silica $d=0.97\pm 0.05\mu m$; (c) S3 silica $d=0.73\pm 0.04\mu m$.}
\end{figure}

\begin{figure}
\includegraphics[height=16cm]{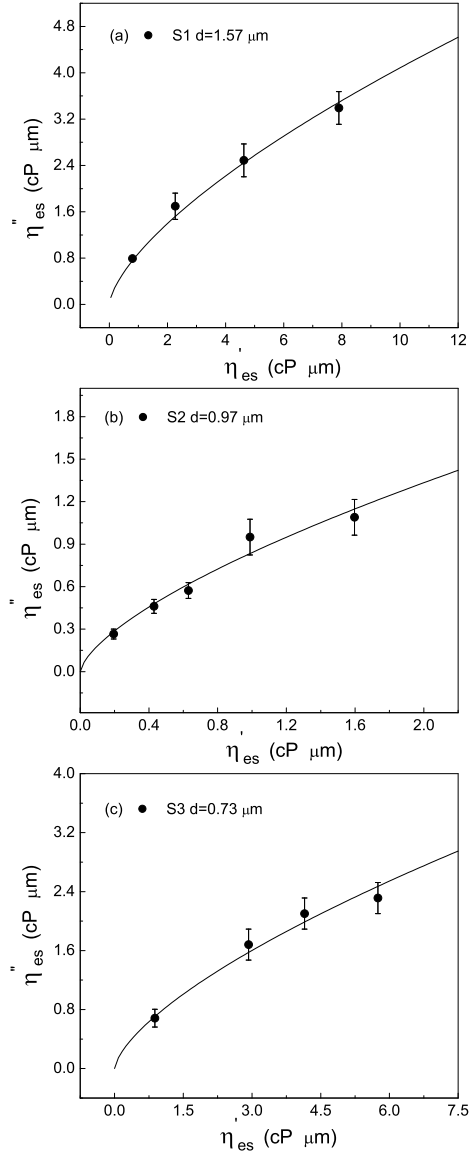}
\caption{\label{fig:epsart} The effective surface viscosity $\eta^{''}_{es}$ versus $\eta^{'}_{es}$ for particles S1, S2 and S3. The samples are the following: (a) S1 silica $d=1.57\pm 0.06\mu m$; (b) S2 silica $d=0.97\pm 0.05\mu m$; (c) S3 silica $d=0.73\pm 0.04\mu m$. The solid lines are the fitted curve $\eta^{''}_{es}=K(\eta^{'}_{es})^{2/3}$, where $K=0.88\pm 0.02$, $0.84\pm 0.04$ and $0.77\pm 0.03$ for (a), (b) and (c), respectively.}
\end{figure}

\begin{figure}
\includegraphics[height=6cm]{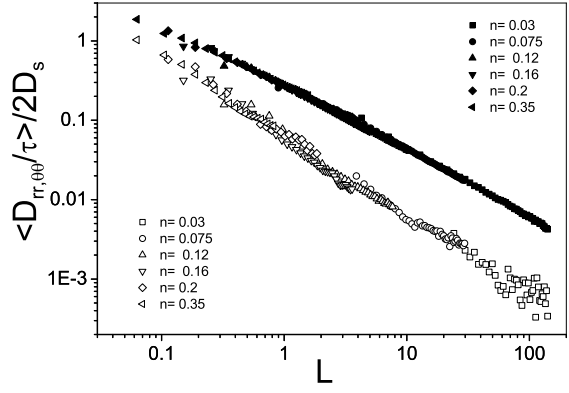}
\caption{\label{fig:epsart} Master curve of the correlation function $<D_{rr}/\tau>$ (solid symbols) and $<D_{\theta\theta}/\tau>$ (open symbols) (scaled by $2D_s$) as a function of reduced particle separation $L$ with various values of area fraction $n$. Different symbols represent the data for different area fraction $n$. The data and symbols in Fig. 6 are same as that in Fig. 2 (a). The samples are S1 silica $d=1.57\pm 0.06\mu m$.}
\end{figure}

\begin{figure}
\includegraphics[height=6cm]{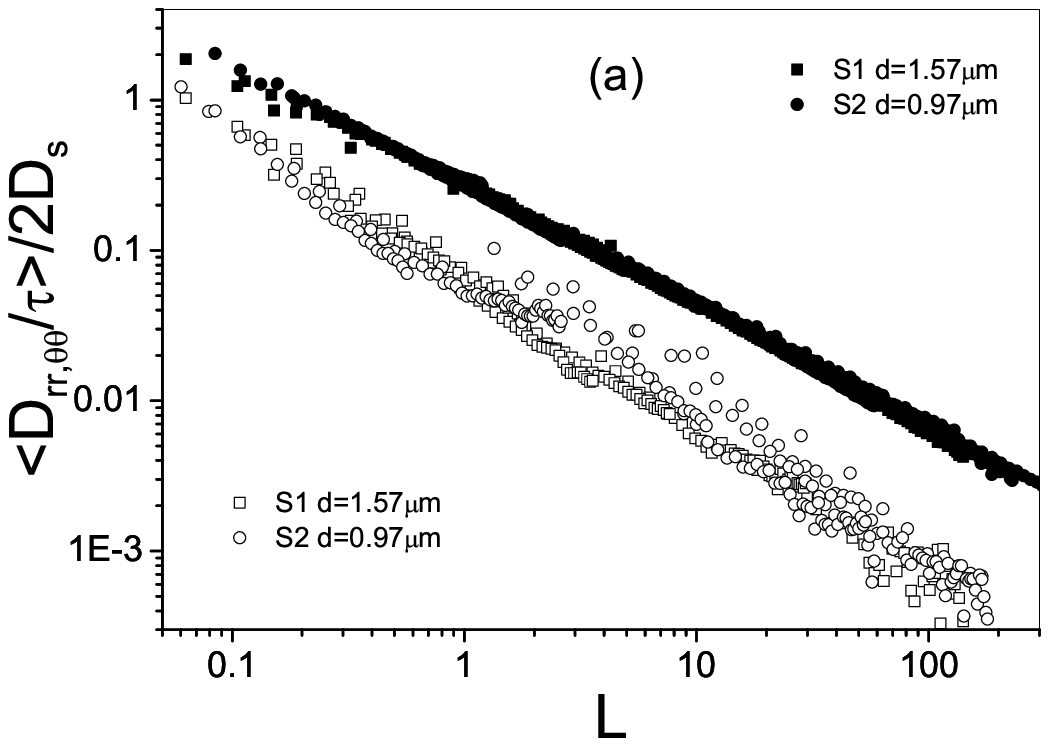}
\includegraphics[height=6cm]{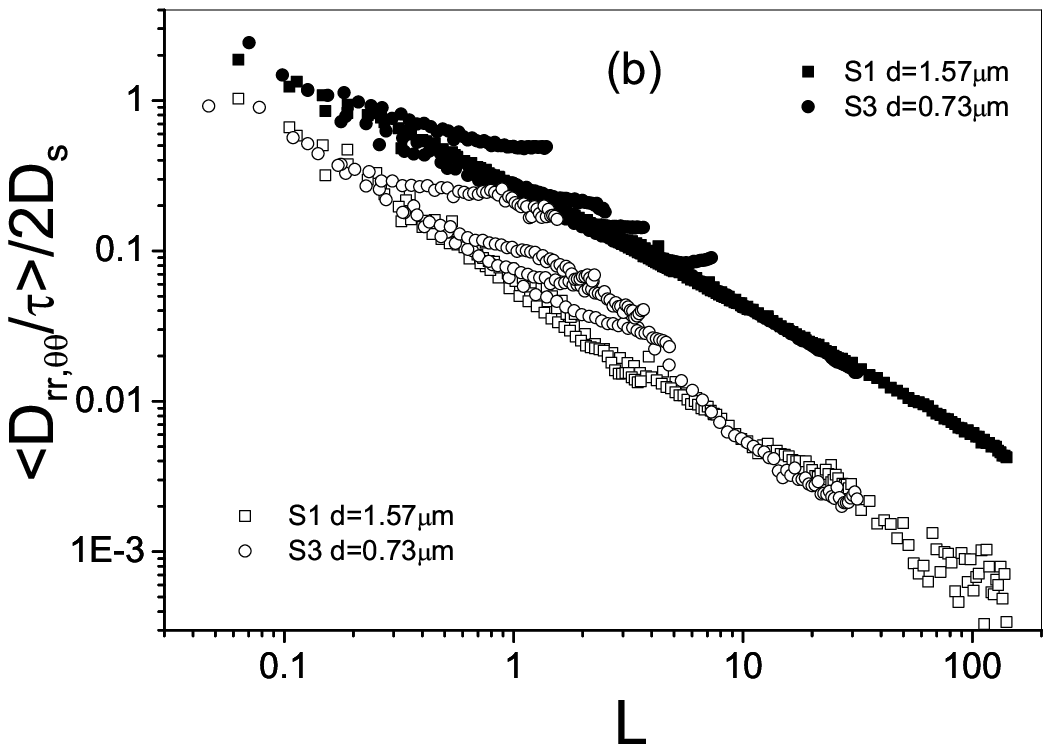}
\caption{\label{fig:epsart}  (a) The mergence of master curve for particles S1 and S2. The scaled variables $\overline{D}_{rr}=<D_{rr}/\tau>/2D_s$ (solid symbols) and $\overline{D}_{\theta\theta}=<D_{\theta\theta}/\tau>/2D_s$ (open symbols) as a function of reduced particle separation $L$ for S1 and S2 particles can collapse onto a single curve. (b) The mergence of master curve for particles S1 and S3. The scaled variables $\overline{D}_{rr}=<D_{rr}/\tau>/2D_s$ (solid symbols) and $\overline{D}_{\theta\theta}=<D_{\theta\theta}/\tau>/2D_s$ (open symbols) as a function of reduced particle separation $L$ for S1 and S3 particles can collapse onto a single curve. Different symbols represent the master curve of different particles.}
\end{figure}

\begin{figure}
\includegraphics[height=16cm]{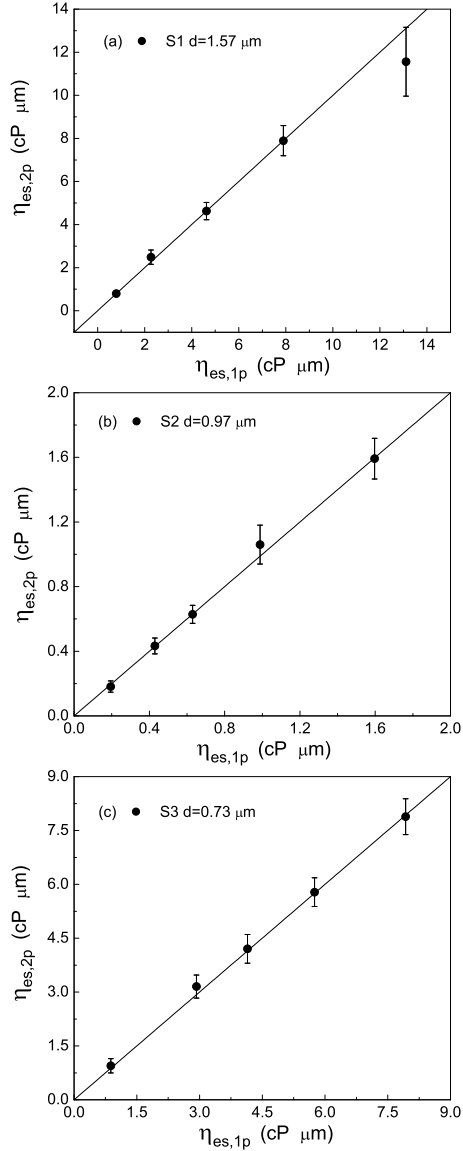}
\caption{\label{fig:epsart} Two-particle surface viscosity $\eta_{es,2p}$ versus one-particle surface viscosity $\eta_{es,1p}$ for particles S1, S2 and S3. The samples are the following: (a) S1 silica $d=1.57\pm 0.06\mu m$; (b) S2 silica $d=0.97\pm 0.05\mu m$; (c) S3 silica $d=0.73\pm 0.04\mu m$. The slope of the solid lines is 1, indicating an equality between $\eta_{es,2p}$ and $\eta_{es,1p}$.}
\end{figure}

\begin{figure}
\includegraphics[height=6cm]{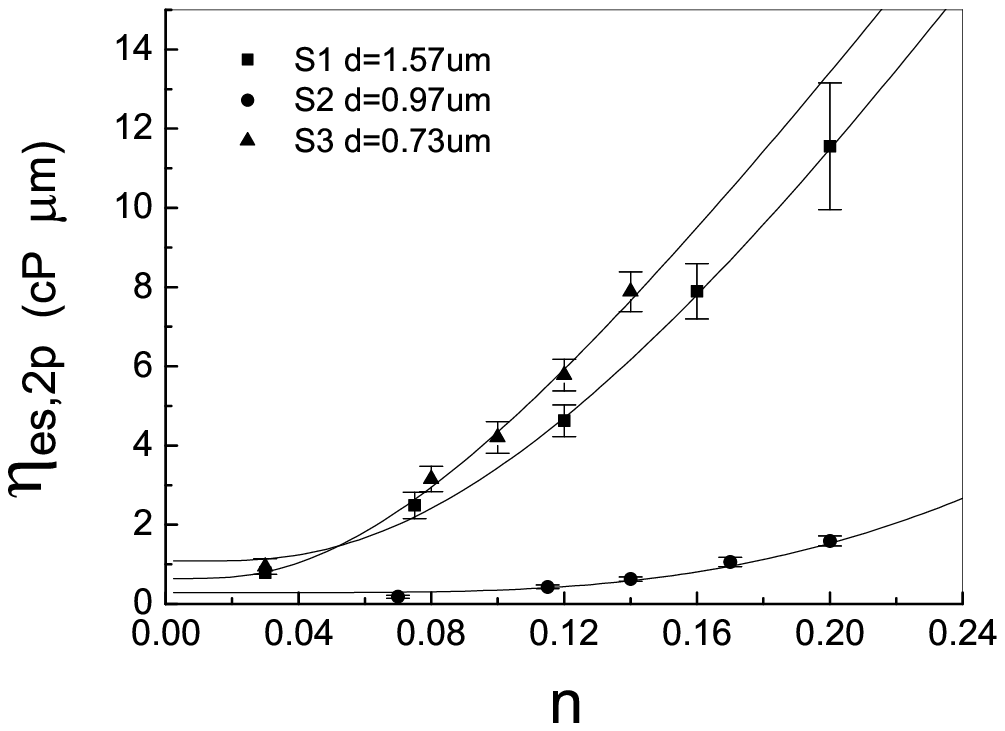}
\caption{\label{fig:epsart} Two-particle surface viscosity $\eta_{es,2p}$ as a function of area fraction $n$ for particles S1, S2 and S3. Different symbols represent the data of different particles. The solid lines are the fitting curve $\eta_{es,2p}=\frac{\kappa} {d} \sqrt{\frac {4n} {\pi}} exp({-\frac {d}{r_0} \sqrt{\frac {\pi} {4n}}})+\eta_{w}r_0$. The fitted values of the parameter $r_0$ and $\kappa$ are presented in table I. }
\end{figure}

Figure 2 (a), 3 (a) and (b) show that $<D_{rr}/\tau>$ and $<D_{\theta\theta}/\tau>$ for various values of particle area fraction $n$ are different. The exponent $\lambda_r$ and $\lambda_{\theta}$ decrease monotonically with the area fraction $n$. Our results shared the similar trend of the work in Prasad $et$ $al.$ \cite{10}, where the exponent is monotonically decreasing by adding HSA protein onto surface. In Prasad's system, the weight of surface viscosity $\eta_{s}$ was modified by HSA. By scaling with $\eta_{s}$, the whole dispersed curves could collapse onto a single master curve. In our data, $D_0$ should be replaced by $D_s=k_BT/4\pi\eta_{es}$, where $\eta_{es}$ is effective surface viscosity. The horizontal axis of R should be scaled by the ratio between the bulk viscosity and the surface viscocity. Following the scaling method of Prasad $et$ $al.$ \cite{10}, we replot the curves for different $n$ by scaling the two sets of effective surface viscosity $\eta^{'}_{es}$ and $\eta^{''}_{es}$ for separation and correlated motion individually. The scaled correlation function and separation are follows:

 \[
\overline{D^{''}}_{rr,\theta\theta}=<D_{rr,\theta\theta}/\tau>/2D^{''}_s, \qquad S=R\eta_w/\eta^{'}_{es},\qquad\qquad (2)
\]

\[
S=R\eta_w/\eta^{'}_{es},\qquad\qquad (3)
\]
where $D^{''}_s=k_BT/4\pi\eta^{''}_{es}$ and $\eta_w$ is the viscosity of the water. We allow them to vary independently to obtain two independent measures of the effective surface viscosity $\eta^{'}_{es}$ and $\eta^{''}_{es}$ as Ref.[10].

Figure 4 (a), (b) and (c) show the relationship of the scaled variables $\overline{D^{''}}_{rr}$, $\overline{D^{''}}_{\theta\theta}$ to the scaled separation $S$ with various values of $n$ for samples S1, S2 and S3, respectively. The data and symbols of Fig. 4 (a), (b) and (c) are same as that in Fig. 2 (a), 3 (a) and (b). The curves with various values of $n$ fall on a single master curve. The curves with higher values of $n$ fall on the left side of the master curve. Figure 5 (a), (b) and (c) describe the scale factors $\eta^{'}_{es}$ and $\eta^{''}_{es}$ used to create the master curve for samples S1, S2 and S3 respectively. As shown in Fig. 5, $\eta^{''}_{es}$ is clearly deviated from $\eta^{'}_{es}$ for the same area fraction. And the discrepancy between $\eta^{''}_{es}$ and $\eta^{'}_{es}$ increases as the area fraction $n$ is creased. The relationship of $\eta^{''}_{es}$ to $\eta^{'}_{es}$ for samples S2 and S3 is qualitatively same as that of particles S1.

The results shown in Fig.5 are hard to understand. The $\eta^{''}_{es}$ and $\eta^{'}_{es}$ should be identical, since they describe the same physics feature. However, experimental results show that the discrepancy between $\eta^{''}_{es}$ and $\eta^{'}_{es}$ is extraordinarily large, e.g., it exceeds 50 percent for samples S2 while $n=0.2$. In Ref.[10], the deviation of $\eta^{''}_{es}$ and $\eta^{'}_{es}$ was only within 15 percent and their averaging values agree well with one-particle viscosity. In our case, the discrepancy is too large to fall in our experimental error range. Especially considering the dependence form of $\eta^{''}_{es}\sim (\eta^{'}_{es})^{2/3}$ in caption of Fig.(5) , the exponent of $2/3$ means that the dimension of  $\eta^{''}_{es}$ on $\eta^{'}_{es}$ are not matching even, which is unreasonable in physics. Checking the above scaling process and comparing with $\eta^{''}_{es}\sim (\eta^{'}_{es})^{2/3}$, we think that the dimension scaling method $\eta_w$/$\eta^{'}_{es}$ may be not necessary way. The fitted index $2/3$ suggests that the following scale method maybe a reasonable choice:

\[
L=\frac R {d}\left(\frac {\eta_{w}d} {\eta_{es}}\right)^{3/2}, \qquad\qquad   (4)
\]
and the correlation function $\overline{D}_{rr,\theta\theta}=<D_{rr,\theta\theta}/\tau>/2D_s$ where $D_s=k_BT/4\pi\eta_{es}$. By this method, we could scale the correlation function and separation onto a single master curve with one set of values of parameter $\eta_{es}$ rather than two sets of parameter value. The obtained value of $\eta_{es}$ by this method is called two-particle effective surface viscosity $\eta_{es,2p}$.

Figure 2(a) for samples S1 was replotted in figure 6 with the new scaling method described in formula (4). The symbols are same as previous. As shown in Fig. 6, all the data sets fall on a single master curve. For samples S2, the behavior of $\overline{D}_{rr}$ , $\overline{D}_{\theta\theta}$ is identical with that of samples S1 and the data sets of the two kinds of particles can fall on the same master curve, as shown in Fig. 7 (a). For particles S3, in the case of short inter-particle distance $R<5d$, the scaled correlated motion $\overline{D}_{rr}$ , $\overline{D}_{\theta\theta}$ for different $n$ can fall on a single master curve, whose behavior is similar to that of particles S1 and S2. However, when $R>5d$ the deviation from the master curve appears, which is different from that of particles S1 and S2. Figure 7 (b) shows that the data of scaled correlated motion $\overline{D}_{rr}$ , $\overline{D}_{\theta\theta}$ of particle S1 and S3 for the case of $R<5d$ can also collapse onto the same master curve. The identicalness of the master curves for different samples suggests that particle size dependence has also been eliminated. The superposition of the data sets for various values of $n$ and different particles indicates that there is a general uniform relationship between cross correlation coefficient $D_{rr}$, $D_{\theta\theta}$ and effective viscosity $\eta_{es}$.

In the experiment, great care is taken to clean the interface to eliminate polymer, surfactant and contamination's effects. We have used Langmuir trough method to measure the changing of surface tension while the area of aqueous surface reduced. No detective impurity was found in the system. We believe that there is no heterogeneity in our experiment system. The two-particle effective surface viscosity $\eta_{es,2p}$ should be nearly equal to the one-particle effective surface viscosity $\eta_{es,1p}$ which is directly related to particles's self-diffusion. To compare $\eta_{es,2p}$ and $\eta_{es,1p}$, we calculate one-particle effective surface viscosity $\eta_{es,1p}$ as follows. At Ref.[14] Chen et al., presented the short-time self-diffusion coefficient $D^s_s$ of silica particles at an air-water interface via the formula $<\Delta r^2(\tau)>=4D^s_s\tau$ for different particle concentration. The measured $D^s_s/D_0$ can be fitted to a second-order polynomial,

\[
 D^s_s/D_0=\alpha(1-\beta n-\gamma n^2),\qquad\qquad  (5)
 \]
where $\alpha$, $\beta$, $\gamma$ are the fitted parameters. At the $n\rightarrow0$ limit, the short-time self-diffusion coefficient $D^s_s(n)$ is directly related to the drag coefficient $\xi$ via the equation $D^s_s(n=0)=k_BT/\xi$. For particles at a distance $z$ from the interface,

\[
\xi=(\eta_1a)f(z/a,B),\qquad\qquad  (6)
\]
where $z$ is the distance between the sphere's north pole and the interface, i.e., $z=0$ when the sphere is in contact with the interface from below, and $a$ is the radius of the sphere \cite{15}. The Boussinesq number $B$ is defined as $B=\eta_s/(\eta_wa)$, where $\eta_s$ is the shear viscosity of the interface and $\eta_w$ is the viscosity of bulk of water. Fischer $et$ $al.$ \cite{15,16} showed that $f(z/a,B)=k^{(0)}+k^{(1)}B+o(B^2)$, which reads the effects of interface viscosity on the single-particle. According to the relations of short-time self-diffusion coefficient to the drag coefficient, the one-particle effective surface viscosity can be written as:

\[
\eta_{es,1p}=\left( \frac{6\pi}{\alpha(1-\beta n-\gamma n^2)}-k^{(0)}\right) \frac{\eta_{w}a}{%
k^{(1)}}.\qquad (7)
\]
The zeroth-order and first-order drag coefficient $k^{(0)}$ and $k^{(1)}$ are given by \cite{15}:

\[
k^{(0)}\approx6\pi\sqrt{\tanh \left( \frac{32\left( z/a+2\right) }{9\pi^2}%
\right) },\qquad\qquad  (8)
\]

\[
k^{\left( 1\right) }\approx\left\{
\begin{array}{c}
-4\ln \left( \frac 2{\pi}\arctan \left( \frac 23\right) \right) \left( \frac{%
a^{3/2}}{\left( z+3a\right) ^{3/2}}\right) \qquad for\quad z>0 \\
-4\ln \left( \frac 2{\pi}\arctan \left( \frac{z+2a}{3a}\right) \right) \qquad
\quad \quad \quad ~for\quad z<0.
\end{array}
\right. \qquad (9)
\]
The zeroth-order drag coefficient $k^{(0)}\approx$ 15.97, 13.0 and 15.97 for samples S1, S2 and S3 respectively \cite{14}. From $k^{(0)}$, we derived the values of $k^{(1)}$ described by formula (9). Substituting $k^{(0)}$, $k^{(1)}$ and the measured values of $\alpha$ $\beta$ and $\gamma$ at Ref.[14] into formula (7), we derived the one-particle effective surface viscosity $\eta_{es,1p}$ for different area fraction $n$.

Figure 8 (a), (b) and (c) exhibit the relationship of the one-particle effective surface viscosity $\eta_{es,1p}$ to the two-particle one $\eta_{es,2p}$ used to create the master curve for particles S1, S2 and S3, respectively. As shown in Fig. 8, the one- and two-particle measurements agree well with each other. The results confirm that the non-heterogeneity argument about our experiment system and the scaling method described in formula (4) are reasonable. But at present we do not derive the scaling relationship of formula (4) theoretically, which deserves further investigation. We vaguely guess the index $3/2$ may be related to the dimension feature of the effective surface viscosity, which dependents on both of the bulk and interface viscosities.

Most interesting phenomenon in the experiment is that the exponent $\lambda_r$ and $\lambda_{\theta}$ decrease when the colloidal area fraction $n$ increases. Experiment results in Fig. 1 indicate that the cross correlation between particles in our experiment is stronger than that exhibited in the experiments of Prasad $et$ $al.$ \cite{10} even in the case of low $\eta_{s}$. By all above analysis, the decrease of $\lambda_r$ and $\lambda_{\theta}$ value is correspond to the increase of the two-particle effective surface viscosity $\eta_{es,2p}$. Figure 9 shows the two-particle effective surface viscosity $\eta_{es,2p}$ as a monotonous increasing function of the area fraction $n$ for samples S1, S2 and S3.  The values of $\eta_{es,2p}$ for particles S2 are extraordinary smaller than that of S1 and S3.

\section{Discussion}

Silica spheres representing an important class of charged particles commonly used in colloidal science have anionic $SiO^-$ groups on their surface \cite{17,18}. These charged spheres at the air-water interface show a DLVO repulsive interaction \cite{14}. The interactions between silica spheres at air-water interface with low fraction $n$ include hydrodynamic and Coulomb interactions. With the increase of $n$, the mean particle separation $\ell=d\sqrt{\pi/(4n)}$ decreases and the mean Coulomb interactions between particles increase. The Coulomb interaction between colloidal particles should serve as an effective surface viscoelastical role in our system and enhance the cross correlation of particles diffusion (shown as Fig.6 and Fig.9). Regarding the effective surface viscosity $\eta_{es,2p}$ as a measurable scale of the energy-cost due to the deformation of interfacial particles configuration, the dependence of $\eta_{es,2p}$ on $n$ should share the feature of interaction potential among the silica sphere.

Particles' interaction potential in our system can be well described by the screened Coulomb potential \cite{14}

\[
U_c(R)\propto \frac {exp[-R/\lambda_D]}{R},\qquad\qquad  (10)
\]
where $R$ is the separation between the particles, and $\lambda_D$ is the screening length. Referring to this potential form and relationship between $R$ and $n$, the effective surface viscosity $\eta_{es,2p}$ could be written as:

\[
\eta_{es,2p}=\frac{\kappa} {d} \sqrt{\frac {4n} {\pi}} exp({-\frac {d}{r_0} \sqrt{\frac {\pi} {4n}}})+\eta_{w}r_0,\qquad\qquad  (11)
\]
where $d$ is the particles' diameter and $n$ is the area fraction. Two fitting parameters are $\kappa$ and $r_0$. The parameter $\kappa$ is related to the strength of the influence of Coulomb interaction on the effective surface viscosity. The parameter $r_0$ is a intrinsic length of the interfacial system. Considering the definition of surface viscosity, $r_0$ maybe related to the length of the line that spheres contact with the interface. As an amplitude parameter, $\kappa$ should be related to the surface charge, immerse depth and diameter of the particle, though no exact expression of $\kappa$ couldn't be given at present. Here, we add an extra term of $\eta_{w}r_0$ as a correction to the theory of equation (7), since these silica spheres, especially samples S1 and S3, experience a larger drag force than the value predicted by T.M.Fisher's theory \cite{14}. This term of $\eta_{w}r_0$ represents the surface viscosity that one particle experiences at the air-water interface when $n\rightarrow 0$.

\begin{table}
\caption{\label{tab:table2}Particle sample used in the experiment and the fitted values of $r_0$ and $\kappa$ from the measured effective surface viscosity $\eta_{es,2p}=\frac{\kappa} {d} \sqrt{\frac {4n} {\pi}} exp({-\frac {d}{r_0} \sqrt{\frac {\pi} {4n}}})+\eta_{w}r_0$. }
\begin{ruledtabular}
\begin{tabular}{cccccccc}
 Sample$/$Manu$.$&$d (\mu m)$ &  $r_0 (\mu m)$ & $\kappa (cP \mu m^2)$\\
\hline
silica 1 (S1)$/$Duke Sci$.$& 1.57$\pm$0.06 & 1.13$\pm$ 0.1 & 510$\pm$84  \\
silica 2 (S2)$/$Bangs Lab& 0.97$\pm$0.05 & 0.30$\pm$ 0.05 & 1500$\pm$800  \\
silica 3 (S3)$/$Duke Sci$.$& 0.73$\pm$0.04 & 0.67$\pm$ 0.1 & 160$\pm$34  \\
\end{tabular}
\end{ruledtabular}
\end{table}

The solid lines in Fig. 9 are the fitted curves described by formula (11). The fitted values of parameters $\kappa$ and $r_0$ are given in table 1. The value of $r_0$ for samples S1 is nearly 2 times larger than that of S3, and $\kappa$ close to 4 times in the error range. These ratio between numbers are consistent well with the size ratio between the particles, considering the underlying physical meaning of $\kappa$ and $r_0$, viz. surface charge and immerse contact line. But for samples S2, the values in table give a different story: a much larger $\kappa$ vs. a much less $r_0$. These two numbers are self consistent at least. More surface charge ($\backsim \kappa$) will give a shorter contact line ($\backsim r_0$), since high charged particles always immerse deeper. As mentioned above, the S2 particles were purchased from Bangs Laboratories, while S1 and S3 particles from Duke Scientific. Both were synthesized in different ways \cite{14}.  The difference in surface chemistry may introduce a different immersion ratio.

At a glance, this S2 immerse-deeper picture is sort of conflict with the fact that the $D^s_s/D_0$ value of S2 ($\backsim 1.5$) is higher than that of S1 and S3 ($\backsim 1.2$) when $n\rightarrow 0$ \cite{14}, viz. the particles S2 experience a relative less drag than S1 and S3 on the air-water interface in the dilute limit. While $n\rightarrow 0$, the effective viscosity $\eta_{e}$ is composed of the bulk viscosity $\eta^{'}_{w}$ and interfacial viscosity $\eta_{es}=\eta_{w}r^{'}_0$. The short self-diffusion constant of particle could be written as:

\[
D^s_s(n\rightarrow 0)=\frac{k_B T} {\eta^{'}_{w}+\eta_{es}}=\frac{k_B T} {\eta_w a k^{(0)}+\eta_w r^{'}_0},\qquad\qquad  (12)
\]
where $a$ is the particle radius, and $r^{'}_0$ is the character length. The deeper the particles immerse, the larger the bulk viscosity $\eta^{'}_{w}$ contributes and the less the $\eta_{es}$. Base on our data, the increase of $\eta^{'}_{w}$ might be overwhelmed by the decrease of $\eta_{es}$ when S2 immerse deeper.

The $r^{'}_0$ in equation (12) should be same with the $r_0$ appeared in equation (11), since both describe the interfacial viscosity $\eta_{es}$ when $n\rightarrow 0$. Regarding $r_0$ as the length of interfacial particle contact line, we could calculate the value of $r^{'}_0$ by substituting the fitting value of $r_0$ (in table I) and the value of $D^s_s$ (in paper \cite{14}) into the formula (8) and (12). The calculation results show that $r^{'}_0=0.97\pm 0.1 \mu m$, $ -0.87\pm 0.1 \mu m$ and $0.47\pm 0.1 \mu m$ for samples S1, S2 and S3, respectively. The deviation between $r_0$ and $r^{'}_0$ is less than 15 and 30 percent for sample S1 and S3, respectively. However, the sign of the $r^{'}_0$ is negative for S2. With a long shot guess, it maybe because that the particle S2 has a partially slip boundary condition due to its unique surface chemistry. Then the fitting value of $r_0$ is not the real length of particles contact line, but a viscosity-effect-equivalent one, which could much less than the real contact length. By this underestimated $r_0$, we could have a overestimated $\eta^{'}_{w}$ due to a deeper-immersion estimation of $z/a$ in equation (8). In the end, we get a minus $r^{'}_0$ for sample S2. But at present, we are unable to experimentally measure the exact length of the interfacial particle contact line independently. So we cannot say anything much more here.

Traditionally, particle concentration dependence of the bulk viscosity (data points in Fig. 9) could be expressed by the polynomial \cite{19,20}. By analogy with the bulk case, we rewrite the effective surface viscosity $\eta_{es,2p}$ as follows:

\[
\eta_{es,2p}=\eta_{es0}\left( 1-n \right)^{-\mu},\qquad\qquad  (13)
\]
where $\eta_{es0}$ and $\mu$ are the fitting parameter. The parameter $\eta_{es0}$ represents the effective surface viscosity that a sphere experiences. The area fraction dependence of $\eta_{es,2p}$ is represented by $\mu$. The data points in Fig. 9 could also fitted very well by equation (13). However, the fitted values of $\mu$ are very high and $\mu\simeq 12\rightarrow 15$, indicating $\eta_{es,2p}$ has a super strong area fraction dependence. Such incredible high power exponent suggests that $\eta_{es,2p}$ should exponentially increases with colloidal area fraction. We believe that the strong area fraction dependence of $\eta_{es,2p}$ does arise from the Coulomb interaction between particles.

Generally, figure 9 exhibits dependence of shear viscosity $\eta_{es,2p}$ on particle concentration, which is directly related to cross-correlation of interfacial diffusion. With the increase of area fraction, the mean particle separation decreases and the electrostatic interaction between particles enhances. The enhancement of the electrostatic interaction causes the increase of correlated strength between particles. The increase of correlated strength corresponds to the increase of the effective surface viscosity $\eta_{es,2p}$. We describe the electrostatic interaction between particles with the effective surface viscosity $\eta_{es,2p}$. There are hydrodynamic and electrostatic interaction between particles in our experimental system. We call this interaction effective hydrodynamic interaction. With the increase of $n$, the effective hydrodynamic interaction enhances mainly arising from the increase of Coulomb interaction between particles. H. Diamant $et$ $al.$ \cite{9,21} had demonstrated that many-body effects do not influence the hydrodynamic coupling at large distance in quasi-two-dimension (Q2D) colloidal system for momentum diffusion does not contribute to the large-distance coupling. We also observed that the hydrodynamic interaction between hard sphere is independent of particle concentration in the two-dimension (2D) system where the momentum leaks to the third dimension \cite{22}.

Another information given by Fig. 9 is size dependence of $\eta_{es,2p}$. First, we study the behavior of the shear viscosity $\eta_{es,2p}$ for particles S1 and S3 both which were purchased from Duke Scientific with identical surface chemistry. It is shown that for the same area fraction the shear viscosity $\eta_{es,2p}$ of small particles is slightly larger than that of large particles, i.e., small particles experience a larger drag than large particles. One possibility is that smaller particles are separated by a smaller distance ($\ell=d\sqrt{\pi/(4n)}$) as compared to the larger particles at the same area fraction $n$, and they may feel more effective hydrodynamic interactions at air-water interface than the larger particles do \cite{23,24,25}. Compared with particles S1 and S3 ,the S2 particles has very small shear viscosity $\eta_{es,2p}$, as mentioned above, which maybe come from the different surface chemistry.

Figure 3 (b) shows an anomalous long-rang correlation. While $R>5d$, the decay of $<D_{rr}/\tau>$ and $<D_{\theta\theta}/\tau>$ bates with increasing $R$ and does not follow power law. At large inter-particle region, correlated motion are weakly dependent on $R$ and are roughly logarithmic (too narrow data range to tell, indeed). The deviation of correlated motion from the master curve in the case of $R>5d$ indicates another rules governed the system or something else? At present, we don't have an answer yet. During the experiment, we use 10-100 image sequences, each contains 100 images, to calculate $D_{rr}$ and $D_{\theta\theta}$, and the result is further averaged over repeated runs (10$-$20 runs). This corresponds to an average over $10^6-10^7$ particles, ensuring that the statistical averaging is adequate. We conduct the measurements at different times with separately prepared air-water interface. The identical experiment equipments and measurement methods are used for all the three particle samples.  Thus, this anomalous long-rang interaction sounds unlikely caused by inaccurate measurements. But this anomalous correlation occurs only at smallest colloidal S3 system. The diffusion constants of S3 particles is relative larger than two others samples due to the smaller size. An improved microscopy system with even higher time and spacial resolution might be necessary to clarify this phenomenon.

\section{SUMMARY}

In this paper, we systematically investigated the spatial correlated motion of weakly charged silica spheres at a pure air-water interface. Three kinds of silica particles are used in the experiment: silica 1 (S1) with the diameter $d=1.57\mu m$ purchased from Duke Scientific, silica 2 (S2) with the diameter $d=0.97\mu m$ purchased from Bangs Laboratories and silica 3 (S3) with the diameter $d=0.73\mu m$ purchased from Duke Scientific. Optical microscopy and multi-particle tracking are used to measure $D_{rr}/\tau$ and $D_{\theta\theta}/\tau$ as a function of the inter-particle distance $R$ with various values of particle concentration for particles S1, S2 and S3. Correlation function $D_{rr}/\tau$ and $D_{\theta\theta}/\tau$ decay with inter-particle distance $R$ as $1/R^{0.86\pm0.02}$ and $1/R^{1.45\pm0.03}$, respectively, for particles S1 with $n=0.03$. With the increase of $n$, the decay of correlated motion with $R$ becomes slow (Fig. 1 and 2). We attribute it as that the cross correlation is enhanced by the Coulomb interaction between particles. We describe the electrostatic interaction between particles at the interface with the effective surface viscosity $\eta_{es}$.

Following the scaling method given by Prasad $et$ $al.$ \cite{10}, we scale the curves with various values of $n$ onto a single master curve by $\eta^{'}_{es}$ and $\eta^{''}_{es}$. The measurements results show that $\eta^{''}_{es}$ is clearly deviated from $\eta^{'}_{es}$ for a certain area fraction (Fig. 5). The discrepancy between $\eta^{'}_{es}$ and $\eta^{''}_{es}$ even exceeds 50 percent in the case of large $n$. The dependence of $\eta^{'}_{es}$ on $\eta^{''}_{es}$ suggests that we should rescale the separation according to formula (4). With this new scaling method the curves with different area fraction can be scaled onto a single master curve by one set of parameter values of two-particle effective viscosity $\eta_{es,2p}$ (Fig. 6). The measurements of effective surface viscosity $\eta_{es,2p}$ and $\eta_{es,1p}$ agree well with each other (Fig. 8). The behaviors of cross correlation for particles S1, S2 and S3 are similar and their master curve can also fall on a single curve except the large $R$ region of S3 (Fig. 7). The experimental results indicate that our new scaling method is reasonable and could be used in similar experiment system. But presently we do not deduce theoretically the scaling relationship described in formula (4), which deserves further investigation, and also cannot understand the anomalous long-rang correlation exhibited by particles S3 at large inter-particle distance region (Fig.3 (b)).

The effective surface viscosity $\eta_{es,2p}$ as a function of $n$ for samples S1, S2 and S3 are presented (Fig. 9). For samples S1 and S3 with identical surface chemistry, $\eta_{es,2p}$ of small particles is slightly larger than that of large one for the same area fraction. However, the viscosity $\eta_{es,2p}$ of particles S2 is extraordinary small compared with that of S1 and S3. The difference of surface chemistry characteristic and immersion depth between particles S1, S3 and S2 may be the two possibilities causing the discrepancy of the viscosity $\eta_{es,2p}$. The experimental data about the effective surface viscosity $\eta_{es,2p}$ presented in the paper are helpful for the related theoretical investigation. Referring to the form of the interaction potential between particles, we present a exponential fitting formula (equation (11)), which is more physically reasonable than the traditional polynomial form, for the viscosity $\eta_{es,2p}$.

The decay of the correlated motion with $R$ for lowest $n$ in this paper is slower than that of Ref.[10] with low surface viscosity. We argue that the high sensitivity of electrostatic interaction between particles on area fraction results in this deviation. With the increase of $n$, the electrostatic interaction enhances and the slope of correlated motion vs. $R$ decreases, which confirm our argument. We increase the Coulomb interaction between particle by increasing the area fraction $n$. Though we have known that many-body effects would not influence the hydrodynamic interaction of particles, the method varying Coulomb interaction between particles in the paper is indirect. We will change the surface charge of particles directly to investigate the effects of Coulomb interaction on correlated motion in the next experiment to further confirm our argument.

\end{document}